\documentclass[aps,pra,twocolumn,groupedaddress,showpacs]{revtex4}

\usepackage{graphicx}
\begin{document}
% You should use BibTeX and apsrev.bst for references
%\bibliographystyle{apsrev}

%\preprint{}

\title{Quantum key distribution using non-classical
\\photon number correlations in macroscopic light pulses}
%\title{Quantum key distribution using macroscopic, non-classical light pulses}
% Optional argument for running titles on pages
%\title[]{}


% other informatio
\author{A.C. Funk and  M.G. Raymer}
%\thanks{}
\affiliation{Oregon Center for Optics and Department of Physics,
University of Oregon, Eugene, Oregon}


\date{\today}

\begin{abstract}
We propose a new scheme for quantum key distribution using
macroscopic non-classical pulses of light having of the order
$10^6$ photons per pulse. Sub-shot-noise quantum correlation
between the two polarization modes in a pulse gives the necessary
sensitivity to eavesdropping that ensures the security of the
protocol.   We consider pulses of two-mode squeezed light
generated by a type-II seeded parametric amplification process.
We analyze the security of the system in terms of the effect of an
eavesdropper on the bit error rates for the legitimate parties in
the key distribution system.   We also consider the effects of
imperfect detectors and lossy channels on the security of the
scheme.
\end{abstract}

%\keywords{}


\maketitle


%\section{Introduction}
Quantum key distribution (QKD) is used for sending a key  from one
party (Alice) to another (Bob) in such a manner that the laws of
quantum mechanics guarantee the security of the key
\cite{ref:bb84}.  The key can be used later as a one-time pad to
encrypt a message.  If an eavesdropper (Eve) intercepts all or
part of the key, errors in the key  are unavoidably introduced,
which are detectable by Bob and Alice, thus revealing the presence
of eavesdropping. \par

All experimentally demonstrated schemes to date have used single
photons to encode the key bits \cite{ref:zbinden}.  Such systems
are subject to several difficulties - foremost is the absence of
reliable technologies for generating single-photon pulses on
demand.  Usually, highly attenuated laser pulses are used to
approximate single-photon pulses, but the presence of the
two-photon component in such pulses provides a potential avenue
for an eavesdropper to foil the security by acquiring the
redundant photon and making measurements on it.  Other
difficulties include sensitivity to stray light and the difficulty
of low-noise detection of single photons at wavelengths (1.3 - 1.5
$\mu m$) that are used in fiber-optic telecommunication.    \par

We present  a scheme for QKD that uses macroscopic, non-classical
light pulses.  Our light pulses are macroscopic in that each
contains on average $10^4-10^6$ photons, and are ``non-classical''
in that their density operator cannot be represented as a
statistical (diagonal) mixture of coherent states.   The variable
that we use for the encoding of each bit is the difference of the
numbers of photons in two optical modes.
\par

Other proposed schemes for QKD using multi-photon non-classical
optical fields exist \cite{ref:ralph2,ref:hillery,ref:preskill}.
These schemes are distinct from ours in that they are based on the
measurement of field quadratures in squeezed states.   These
schemes require one or more local oscillators, that are phase
locked to the signal field, for the measurement of the signal
field, thereby introducing a practical difficulty in the
implementation.  Schemes based on polarization require not phase
stability, but polarization axis stability.  \par

There exists a proposal to use macroscopic optical pulses prepared
in a coherent state, to perform QKD\cite{ref:bennet_patent}.
This scheme uses the inherent quantum uncertainty for the number
of photons in the coherent state to ensure security.  However the
intended recipient of the key is also subject to the same
uncertainties in photon number as the eavesdropper.  This results
in a large systematic error rate ($\geq 30\%$) for the measurement
of the bits.  The large error rate is corrected by using
considerable amounts of classical error correction,    requiring a
large number of optical pulses to be sent for each logical key
bit.   This results in a very low logical-bit-per-optical-pulse
rate.    \par

In this paper we will describe how our scheme overcomes some of
the difficulties with practical implementations for previously
proposed QKD schemes. We also provide a strong plausibility
argument for the security of our scheme. \par

It should be noted that this paper does not contain a proof of
absolute security, as has been proven for the BB84 protocol
\cite{ref:mayers} and the quadrature squeezed-state protocol
\cite{ref:preskill}.  The first proof of security for BB84
\cite{ref:mayers} did not take into account several practical
points.  Subsequently, several proofs have been constructed which
take into account some important practical considerations which
affect the security of QKD protocols such as lossy channels,
imperfect detectors and imperfect sources
\cite{ref:mayers3,ref:mayers2,ref:lo_chau,ref:preskill2}. Proof of
absolute security is a difficult task, and work is currently
underway to construct a rigorous proof of absolute security at
non-zero data rates for our QKD protocol.
\par

%\section{Security Overview}

In our protocol the security of the key is ensured by using
non-classical light pulses having, in a particular polarization
basis, a photon difference number between two polarization modes
that is better defined than in a coherent-state with the same
total number of photons. The quantum correlations between the
orthogonal polarization modes are rapidly degraded by any of Eve's
attempts to measure the key.  The degradation of the correlations
leads to different measurement results for Bob than in the
eavesdropper-free case.   The changes in the measurement result
will then indicate to Bob and Alice the presence of Eve. \par

%\section{Protocol overview}

The protocol is as follows: Alice encodes each bit value in the
mean ``polarization difference number'' $\langle n\rangle=\langle
n_1-n_2\rangle$ , where $\langle n_1\rangle(\langle n_2\rangle)$
is the mean number of photons in the first (second) polarization
mode making up a basis. Two different polarization bases are used.
One basis (``V/H basis'') is defined by  the vertical and
horizontal linear polarizations; then $ n= n_V-n_H$.  The other
basis (``$\pm45$ basis'') is  defined by the +45 degree linear
polarization and the -45 degree linear polarization; then $
n=n_{+45}-n_{-45}$.  Alice chooses at random which basis to use.
Bob measures the photon difference number either in the V/H basis
or in the $\pm$45 basis.  After the transmission of all the key
bits, Alice and Bob communicate via a public channel and compare
which basis was used on each encoding/measurement.  Alice and Bob
will keep only the bits for which they used the same basis for the
respective pulse. To estimate the overall error rate, Bob and
Alice compare a small fraction of the key bits over the public
channel.   \par

%\section{Generation of Pulses}

We consider non-classical optical pulses used for the encoding
generated using a type-II seeded parametric amplification process
\cite{ref:heidemann, ref:aytur,ref:smithey}.  The amplifier
consists of a type-II non-linear optical crystal which is pumped
by a vertically polarized optical pulse at frequency $2\omega$.
The crystal is simultaneously seeded with a transform-limited
optical pulse at frequency $\omega$.  Each polarization mode of
the  seed pulse is in an independent  coherent state, with a mean
number of photons $\langle n_V\rangle $ in the vertical
polarization mode and a mean number $\langle n_H\rangle$ in the
horizontal polarization mode. Both the vertical and horizontal
polarization modes at $\omega$ experience amplification.   The
overall amplification can be characterized by $G(>1)$, which is
the factor by which the total mean photon number $N_{T}=\langle
n_{V} + n_{H} \rangle$ increases.  $G$ values of up to 20 have
been experimentally measured  \cite{ref:smithey}. The mean photon
difference number $\langle n\rangle=\langle n_V-n_H\rangle$ of the
seed pulse is small ($\leq 1 \%$) compared to  $\langle n_V\rangle
$ and $\langle n_H\rangle$.    \par

%\section{Amplification}

Quantum correlations between the vertical and horizontal
polarization modes are generated by the amplifier, which result in
the statistical properties (including the mean and the variance)
of the photon difference number $n$ to remain unchanged by
amplification.    This follows from the fact that $n$ is a
conserved quantity under the action of the nondegenerate two-mode
squeezing (parametric amplification) Hamiltonian, which produces a
non-classical state of light.\cite{ref:mollow} \par

%\section{Pulse properties}

For coherent-state seed pulses, the variance of $n$ equals the
total mean number of photons in the seed pulse,
$\textrm{var}(n)_{seed}=\langle n_V + n_H\rangle_{seed}$.  The
variance of $n$ for the amplified pulse is the same as the
variance of $n$ for the seed pulse, therefore,

\begin{eqnarray}
\textrm{var}(n)_{amp}&=\langle n_V  +  n_H \rangle_{seed}
&=\frac{1}{G} N_{T,amp}
\end{eqnarray}

The variance of $n$ after the parametric amplification is thus
considerably smaller than the variance that would be present if
the amplified pulse were in a coherent state having the same $N_T$
as in the amplified pulse.  For the coherent-state case, the
variance would be given by the total number of photons,
$\textrm{var}(n)_{coherent}=\langle n_V +  n_H
\rangle_{amp}=N_{T,amp}$.   This coherent-state variance is
referred to as the shot-noise level (SNL).  Thus the variance of
$n$ for the time-integrated amplified pulse will be below the SNL
by a factor of $G$ compared to a coherent-state  pulse with the
same number of photons \cite{ref:smithey, ref:mollow}.  The extent
to which the variance of $n$ is below the SNL, tells us how strong
the quantum correlations between the photons in the vertical and
horizontal polarization modes are.\par

%\section{The different Bases}

The phase difference between the vertical and horizontal
polarization modes is $\pi /2$, giving a polarization state that
is very nearly circular, with a  slight degree of  ellipticity
determined by $\langle n\rangle$.  The major axis of the
polarization ellipse is oriented vertically in the case where
$\langle n\rangle>0$  and oriented horizontally in the case where
$\langle n\rangle<0$.  Alice can switching the bit value by
performing a 90 degree rotation of the slightly elliptical
polarization state. \par

Following the parametric amplifier is a 45 degree polarization
rotator. Alice can use this to rotate the polarization by 45
degrees before sending the pulse to Bob.  This will have the
effect of changing the V/H basis into the $\pm$45 basis.  This
polarization rotation  is applied or not at random.   Alice
records which basis (V/H or $\pm$45) was used for each pulse.  For
those pulses that have their polarization basis rotated, the bit
encoding changes.   The relevant mean photon difference number
$\langle n\rangle$ is now written as $\langle n\rangle=\langle
n_{+45}-n_{-45}\rangle$.  We will refer to the basis that is set
by Alice's rotator on a given pulse as the ``correct'' basis  and
the other basis as the ``incorrect'' basis. \par

%\section{Bob's Measurements}

Bob receives the optical pulses sent by Alice.  Bob measures $n$
in either the V/H basis or in the $\pm 45$ basis at random.   Bob
uses a 45 degree polarization rotator and a polarizing beam
splitter to select a basis and separate the  polarization modes.
He counts the number of photons in each of the polarization modes
for a given basis (within precision set by detector noise), and
subtracts the number of photons in each mode to determine  $n$.
\par

Alice encodes a logical ``1'' (``0'') key bit by setting the mean
value of the difference number to be in the correct basis $\langle
n \rangle = + N \,(\langle n \rangle =-N)$, where $N$ is a
positive number comparable to  $Sqrt{N_{T,amp}}$, the SNL for the
total field.\par

The action of the basis change on the two-mode photon-correlated
state produced by the OPA results in two independent single mode
quadrature squeezed states in the polarization modes of the
incorrect basis.  There are no correlations between these
quadrature squeezed states.    Therefore, $\langle n\rangle=0$,
regardless of the bit value, and the variance of $n$ in the
incorrect basis is thus the variance of $N_{T,amp}$, which can
easily be calculated from equations \ref{eq:beamsplitter}.  This
variance is always greater than $N_{T,amp}$.    There is thus
greater uncertainty for a measurement of $n$ in the incorrect
basis  than in a coherent state with $N_{T,amp}$ photons.  \par

\begin{figure}
\includegraphics[width=8.6cm,height=6.5cm]{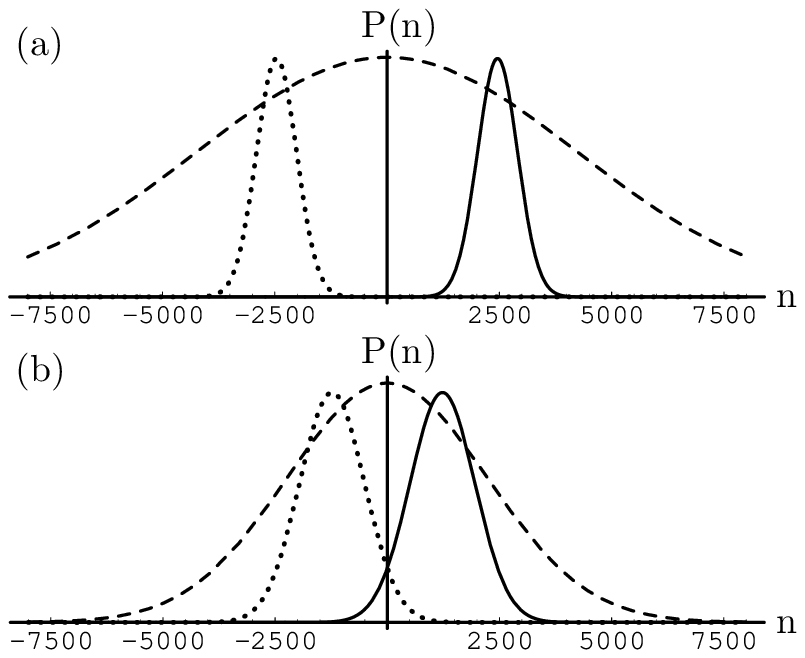}
\caption{Probability distributions for measurement of $n$ in the
correct basis for a logical 1 (solid curve), logical 0 (dotted
curve), and for measurement in the incorrect basis (dashed curve)
for (a) no loss and (b) for 50\% loss} \label{fig:prob_dist}
\end{figure}

By setting, in the correct basis, $|\langle
n\rangle|=N<<N_{T,amp}$, a single measurement of $n$, regardless
of whether the measurement was made in the correct or incorrect
basis, will result in a numerical value within the same range.
This can easily be seen from the distributions for measurements of
$n$ shown in Figure \ref{fig:prob_dist}.   This makes it difficult
to determine from a single measurement which basis is  correct and
which is incorrect.
\par

Bob  decodes a measurement yielding $n>0 $ as a logical ``1'', and
$n<0$ as a logical ``0''.  Bob does not know {\it{a priori}} which
basis Alice used to encode each key bit.  In the incorrect basis,
the probability distribution $P(n)$ for the photon difference
number $n$ is the same regardless of the bit value Alice sent.
There is thus no key bit information contained in the results of a
measurement in the incorrect basis.    To eliminate the results of
such measurements, after the transmission of all the bits, Bob and
Alice communicate publicly to determine on which pulses Bob was
using the correct basis.  The bits are kept only for those pulses
for which Bob was measuring in the correct basis.  \par

Bob does not need to use an ``ideal'' detector to measure the
number of photons in each of the polarization modes.  Due to the
finite width of the initial Poisson distribution for the photon
number in the coherent seed, it is not necessary to use a detector
that can distinguish between $m$ and $m\pm1$ photons.  In practice
a detector with a noise-equivalent photon number around 200-300 is
sufficient.  This allows the use of standard linear photodiodes
with quantum efficiencies approaching 100\% \cite{ref:smithey}.
Even non-unity quantum efficiency detectors are acceptable, with
deviation from unity efficiency simply treated as a loss, which
will be discussed below.\par

Using $\hat{n}=\hat{n}_V-\hat{n}_H={\hat{a}_{V}}^{\dagger}
\hat{a}_{V} - \hat{a}^\dagger _H \hat{a}_H$, where the $\hat{a}$'s
are boson annihilation operators, we calculate the moments of $n$
by writing the amplified annihilation operators in terms of the
seed annihilation operators and assuming coherent-state seed
pulses.   The annihilation operators for the amplified pulses when
they reach Bob, including any losses experienced by the pulse
during the propagation, are given by  the two-mode squeezing
transformation combined with a non-polarizing linear beamsplitter
transformation to account for the losses\cite{ref:smithey}
\begin{eqnarray}
\hat{a}_{V(a)}&=& \sqrt{1-\eta}\left(\mu \hat{a}_{V(s)} + nu
\hat{a}_{H(s)}^\dag\right) +
i\sqrt{\eta}\,\hat{b}_V\\
\hat{a}_{H(a)}&=& \sqrt{1-\eta}\left(\mu \hat{a}_{H(s)} + \nu
\hat{a}_{V(s)}^\dag\right) + i\sqrt{\eta}\,\hat{b}_H,
\label{eq:beamsplitter}
\end{eqnarray}
where the $(a)$ subscripts refer to the amplified pulse, the $(s)$
subscripts to the seed pulse, and $\eta$ is the loss experienced
by the pulse during propagation.   The loss parameter $\eta$
includes loss due to a lossy transmission channel and loss due to
partial sampling of the beam by an eavesdropper.  The $\hat{b}$'s
are the boson operators for the vertical and horizontal vacuum
modes associated with the losses, and $\mu$ and $\nu$ are complex
non-linear coefficients obeying $|\mu|^2 - |\nu|^2=1$, which are
functions of the properties of
the non-linear crystal and the pump beam. \par

By calculating the appropriate moments of $\hat{n}$, we can get
the probability distributions for $n$. Shown in Fig.
\ref{fig:prob_dist}(a) are the unnormalized probability
distributions for Bob's measurement of $n$ in the correct basis
with a sent logical 1 key bit (solid curve),   with a logical 0
key bit (dotted curve), and for a measurement in the incorrect
basis (this distribution is the same for both logical 1 and 0 key
bits) (dashed curve).   For Fig. \ref{fig:prob_dist}(a), in the
case of 100\% transmission efficiency, the following realistic
numerical values were used: $\pm N=\pm 2460$,  $\mu=1.7$, and
$\nu=2.36 e^{i \pi/2}$, leading to $G=10$ and $N_T=2 \times 10^6$
after the amplification.   These parameters lead to a variance of
$n$ in the correct basis that is 10 times smaller than
the SNL, when there is no loss (i.e. $\eta=0$). \par

The distributions plotted in Fig. \ref{fig:prob_dist} for
measurements made in the correct basis are Gaussian approximations
of the Poisson distributions for $n$, with means and widths
determined by the calculated means and variances of $n$.     The
Poisson distributions are very well approximated (to better than
$10^{-5}$) by Gaussian distributions for pulses with photon
numbers $>10^4$.  \par

The distributions plotted in Fig. \ref{fig:prob_dist} for
measurements made in the incorrect basis are Gaussian
approximations of the exact distribution.  In the wrong basis,
each polarization mode is in an independent single-mode
quadrature-squeezed state. The photon number distribution for each
single-mode squeezed state can in our limit of large photon number
be well approximated by a Gaussian distribution
\cite{ref:loudon_knight}.  The difference of two independent
Gaussian variables will thus also be Gaussian.
\par

Due to the tails of the distributions for the two bit values,
there is a non-zero probability that a pulse encoded by Alice as a
logical 1(0) would be measured as a logical 0(1).  Such an error
is a ``bit-flip error'' (i.e.  $1\rightleftharpoons 0$).  Using
the same numerical values for the system parameters, the error
rate in the absence of loss or an eavesdropper is $10^{-8}$. \par

One of the effects of losses or an eavesdropper is to increase
Bob's error rate in a noticeable way.  The change in the error
rate due to the eavesdropper depends on the particular type of
attack and the extent of the attack.  It should be noted that
there exist situations (such as the ``superior-channel attack''
discussed below) where the eavesdropper can take advantage of
large losses to acquire key information.
\par

\begin{figure}
\includegraphics[width=8.6cm,height=4.5cm]{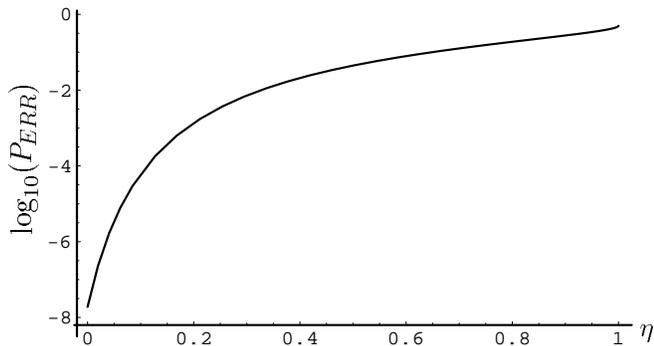}
\caption{Bob's and Alice's error rate $P_{ERR}$ versus  loss
$\eta$.} \label{fig:error_rate} \end{figure}

Any transmission loss experienced by the pulse will increase the
error rates even in the absence of an  eavesdropper.  Bob and
Alice can determine their systematic error rate  by characterizing
the loss of the transmission medium using classical means before
the QKD system is installed.   Shown in Fig.  \ref{fig:error_rate}
is a $\log$ plot of Bob's (and Alice's) error rate versus the loss
$\eta$.  Any increases from their new systematic error rate will
be indicative of the presence of an eavesdropper.  \par

%\section{All about Eve }

We will analyze four different attacks by Eve on the QKD system.
In the first attack, Eve captures the entire optical pulse sent by
Alice, makes a measurement in a randomly chosen basis, and records
the inferred bit value.  She then attempts to prepare the same
state that Alice sent, and sends the prepared state on to Bob. Eve
does not know which basis was used to encode the bit, and in the
cases  where she measures $n$ in the incorrect basis (50\% of the
pulses), she will get the wrong bit value 50\% of the time. Eve's
errors will result in Alice and Bob having a 25\% bit-flip error
rate, which is a clear indication of eavesdropping.   \par

\begin{figure} \includegraphics[width=8.6cm,
height=4.5cm]{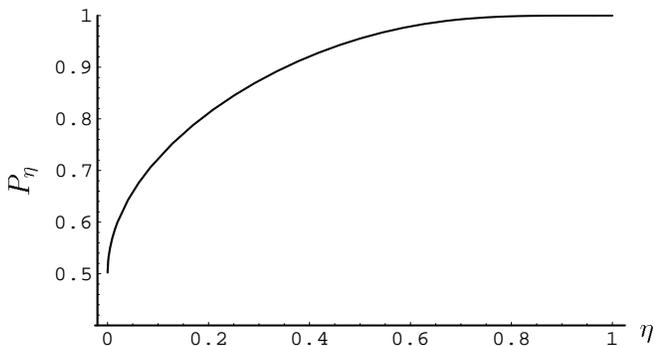} \caption{Eve's probability $P_{\eta}$
to infer the correct bit value given sampling fraction of $\eta$,
versus $\eta$, assuming she correctly guesses the basis.}
\label{fig:eve_info} \end{figure}

In the second attack, Eve simply samples a fraction of the pulse
with a non-polarizing beam splitter and lets the remainder
continue on to Bob.  Eve can then do any sort of measurement on
the sampled portion and try to determine some information about
the key bit.  Any attempts by Eve to sample part of the beam will
result in a loss $\eta$.   As discussed earlier, and as can be
seen from the plot in Figure \ref{fig:error_rate}, there is an
increase in the error rate as $\eta$ increases  from 0.  Based on
their error rate, Bob and Alice can make a good estimate for an
upper bound on the amount of information that Eve would be able to
obtain by sampling with a beam splitter.  \par

In the third attack, Eve captures the entire optical pulse sent by
Alice, passes it through a non-polarizing 50/50 beamsplitter and
measures the photon difference number simultaneously in both bases
(V/H and $\pm 45$).  Based on the results of this measurement, Eve
prepares the state she believes Alice sent, and sends that state
to Bob.  The probability distributions for the difference number
$n$ that Eve would measure in this case are shown in Fig.
\ref{fig:prob_dist}(b).  Given the considerable overlap between
the three possible distributions,  Eve does not gain much
information from the results of a single measurement on both bases
about which
basis was used to encode the bit. \par

Figure \ref{fig:eve_info} shows Eve's probability of inferring the
correct bit value as a function of $\eta$ for sampling a fraction
of the pulse with a non-polarizing beam-splitter.   This plot
assumes that Eve knows which basis is being used, which will in
general not be true, further reducing her knowledge of the key.
In the case of simultaneous measurements, Eve measures at 50\%
sampling.  From Fig. \ref{fig:eve_info}, Eve will have
approximately a 95\% probability of getting the bit value correct
for the correct basis, but she has only approximately a 50\%
probability of getting the basis correct.   Therefore she has only
a 50\% probability of getting the correct bit value and basis.
She will thus prepare states which result in incorrect bit values
for Bob.  These errors will be detected during Bob's and Alice's
error rate checking, once again giving a clear indication of the
eavesdropping.  \par

The fourth attack is a ``superior channel attack''.  It requires
that Eve possess the following technical items:  a quantum memory
system which can store quantum states for a potentially long
period of time and a transmission channel which is lossless. The
attack consists of Eve splitting the optical pulse into two equal
parts using a 50/50 beamsplitter, sending one half of the pulse to
Bob, and keeping the other half of the pulse.  The pulse that she
sends to Bob is sent on Eve's lossless transmission channel which
she has substituted for the original lossy channel.  Eve stores
the states of all the optical pulses sent from Alice to Bob with
her quantum memory system.  Eve then waits until after the public
discussion which reveals the measurement bases, and she then
measures her stored pulses in the correct bases.
\par

In the case that the transmission loss from Alice to Bob (before
Eve replaces the channel with her lossless channel) was 50\%, Bob
and Eve will receive the same information.  Both Eve and Bob will
have received the same optical pulse which has experienced a 50\%
loss. Any one-way error correction which is sent by Alice will
help correct Eve's errors just as well as it corrects Bob's
errors.  In the event that the original transmission loss is
greater than 50\%, Eve will be able to obtain even more key
information than Bob. \par

Such an attack could be avoided by limiting the use of the
protocol to channels with less than 50\% loss.  Also, if the error
correction or privacy amplification required two-way communication
between the recipient of the key, Eve would not necessarily be
able to correct her errors without revealing her identity.  \par

%\section{Error rates}

%\section{Conclusion and Parting Shots}

It is possible to generate type-II parametrically amplified pulses
as described in this paper using conventional lasers and
non-linear crystals.  It is possible to make direct photodetection
measurements of the signal pulses with the necessary
sensitivity\cite{ref:smithey}.  This differs with the QKD schemes
using quadrature-squeezed states which require homodyne detection.
Our scheme also has a low systematic bit error rate, unlike the
coherent-state key distribution system\cite{ref:bennet_patent},
which requires considerable redundancy to overcome intrinsic
uncertainties that are unavoidable for the {\emph{intended}}
recipient of the key.  \par

The physical origin of the security for our QKD scheme lies in the
behavior of non-classical quantum fields when subject to
beam-splitting losses or to polarization-basis changes.  The
plausibility of the security is based on the fact that Eve's
attacks will consist of combinations of beam-splitting and
polarization basis changes.  Other more general attacks need to be
considered further.      \par

\begin{acknowledgments}
We thank the referee for helpful comments about the superior
channel attack.  This material is based upon work supported by the
National Science Foundation under Grant No. 9876608.  \par
\end{acknowledgments}


\end{document}